\documentclass[%
 aip,
 amsmath,amssymb,
 reprint,%
]{revtex4-1}

\usepackage{graphicx}
\usepackage{dcolumn}
\usepackage{bm}

\usepackage{color}
\usepackage[utf8]{inputenc}
\usepackage[T1]{fontenc}
\usepackage{mathptmx}
\usepackage{etoolbox}

\makeatletter
\def\@email#1#2{%
 \endgroup
 \patchcmd{\titleblock@produce}
  {\frontmatter@RRAPformat}
  {\frontmatter@RRAPformat{\produce@RRAP{*#1\href{mailto:#2}{#2}}}\frontmatter@RRAPformat}
  {}{}
}%
\makeatother

\begin{document}

\preprint{APS/123-QED}

\title{Thin-film magnomechanics in the low gigahertz regime}

\author{Yu Jiang}
\affiliation{ 
    Department of Electrical and Computer Engineering, Northeastern University, Boston, MA 02115, USA
}

\author{Zixin Yan}
\affiliation{ 
    Department of Electrical and Computer Engineering, Northeastern University, Boston, MA 02115, USA
}

\author{Yizhong Huang}
\affiliation{ 
    Department of Electrical and Computer Engineering, Northeastern University, Boston, MA 02115, USA
}

\author{Xufeng Zhang$^*$}
\email{xu.zhang@northeastern.edu}
\affiliation{ 
    Department of Electrical and Computer Engineering, Northeastern University, Boston, MA 02115, USA
}
\affiliation{ 
    Department of Physics, Northeastern University, Boston, MA 02115, USA
}

\date{\today}

\begin{abstract}
The coherent interaction between magnons and phonons in the low-GHz regime represents an unexplored frontier in hybrid magnonics, critical for quantum information processing and microwave-to-acoustic transduction. While previous studies have focused on higher frequencies ($>$5~GHz), we demonstrate magnon--phonon coupling near 2~GHz using spoof surface plasmon polariton (SSPP) waveguides integrated with yttrium iron garnet (YIG) thin films of varying thicknesses. SSPP waveguides provide strong slow-wave enhancement, enabling efficient magnon readout in this challenging regime. Systematic measurements reveal the dependence of coupling strength on YIG thickness and phonon wavelength matching, achieving cooperativity $C=1$ for a 3~$\mu$m film at 2~GHz. Angle-dependent studies uncover coupling to both transverse and longitudinal phonon modes. Further investigation shows that thicker films exhibit rich multimode dynamics between high-order magnons and HBAR phonons. These results establish a robust low-GHz magnomechanical platform, opening pathways for multimode quantum transduction and hardware-efficient quantum technologies.
\end{abstract}


\maketitle

\section{Introduction} 
Hybrid magnonics has emerged as a versatile platform for exploring coherent interactions between spin excitations and other information carriers, offering a rich landscape for both theoretical and experimental advancements \cite{Rameshti_PhysRep_2022,Harder_SSC_2018,Lachance_APE_2019,Bhoi_SSP_2020,YiLi_JAP_2020,Awschalom_IEEETransQuantEng_2021,Zhang2023Sep_MTE}. In these systems, magnons---quantized spin waves in magnetic materials---can couple strongly with photons, phonons, and other excitations, forming hybrid states such as magnon polaritons \cite{Huebl_PRL_2013,XufengZhang_PRL_2014,Tabuchi_PRL_2014,Goryachev_PRAppl_2014,LihuiBai_PRL_2015,JustinHou_PRL_2019,YiLi_PRL_2022,2024_PRL_Xu_slowwave} and magnon polarons \cite{XufengZhang_SciAdv_2016,Potts_PRX_2021,2021_PRAppl_Xu_pulseEcho,An_PRB_2020,Hwang_PRL_2024,Shen_PRL_2022,Li2021Jun_APLM,Hatanaka_PRAppl_2022,An_PRX_2022}. These hybrid modes inherit and combine the distinct properties of their constituents, enabling novel functionalities such as nonreciprocal signal transmission \cite{XufengZhang_PRAppl_2020,YipuWang_PRL_2019} and topological energy transport \cite{Lee2023TopologicalMagnonPhoton,HanFast2024PRApplied,Zhang2025TopologicalModeSwitch}. The use of low-damping magnetic materials such as yttrium iron garnet (YIG) facilitates the realization of strong coupling conditions, where coherent energy exchange surpasses dissipation, paving the way for applications in quantum information processing \cite{Tabuchi_science_2015Jul,Lachance-Quirion_Science_2020Jan,Lachance-Quirion2017Jul_SciAdv,Bolski_PRL_2020,Xu_PRL_2023May}, transduction \cite{Hisatomi_PRB_2016,NaZhu_Optica_2020,XufengZhang_PRL_2016}, and fundamental physics explorations such as dark matter detection \cite{ Flower_PhysDarkUniv_2019,Crescini_PRL_2020}.

Within hybrid magnonic systems, magnomechanics---where magnonic and phononic modes are coherently coupled---stands out for its potential in precision measurement and quantum thermodynamics \cite{XufengZhang_SciAdv_2016,Potts_PRAppl_2020,Potts_PRX_2021}. Recent studies have shown that YIG thin-film structures are particularly advantageous for hybrid magnomechanical devices, as magnons in the YIG film can interact with the high-overtone bulk acoustic resonator (HBAR) formed by the YIG/GGG substrate \cite{2021_PRAppl_Xu_pulseEcho,An_PRB_2020,An2023Jul,An_PRX_2022}. These systems have demonstrated novel functionalities such as coherent pulse echo \cite{2021_PRAppl_Xu_pulseEcho}, long-range transfer \cite{An_PRB_2020}, and dark states \cite{An_PRX_2022}. However, most prior demonstrations focus on frequencies above 5 GHz, where the electromagnetic (EM) wavelength is relatively short, resulting in strong microwave photon--magnon coupling and robust excitation and readout signals. In contrast, exploration of magnon--phonon interactions in the low-GHz range---around 2--3 GHz, an important band for quantum systems such as solid-state spin qubits (e.g., nitrogen-vacancy centers in diamonds)---remains limited. At these frequencies, the EM wavelength is much longer, creating a significant volume mismatch between the magnonic and microwave photonic modes. Consequently, the magnon--microwave photon coupling becomes much weaker unless the YIG device footprint is substantially increased, which is undesirable for integrated device development.

In this work, we investigate magnon--phonon interactions in the low-GHz regime. By employing a corrugated microstrip waveguide that supports spoof surface plasmon polaritons (SSPPs), we achieve efficient broadband readout of magnon signals in YIG thin films at around 2 GHz through the slow-wave enhancement \cite{2024_PRL_Xu_slowwave}. This enables systematic exploration of magnomechanical coupling in this previously underexplored frequency band. Strong magnon--phonon coupling is demonstrated using a 3~$\mu$m YIG thin film, and the dependence of the coupling strength on the bias magnetic field orientation is characterized. In addition to transverse phonon modes, longitudinal phonons also emerge when the bias field is tilted at specific angles. Furthermore, multimode strong coupling between multiple phonon and magnon modes is observed, offering opportunities for complex functionalities such as magnon frequency comb generation and multimode magnomechanics based quantum machine learning. Our results provide new insights into magnomechanics in the low-GHz regime and highlight its potential for hybrid magnonic applications.

\section{Hybrid SSPP-magnon-phononic device}
\begin{figure}[tb]
    \centering
    \includegraphics[width=\linewidth]{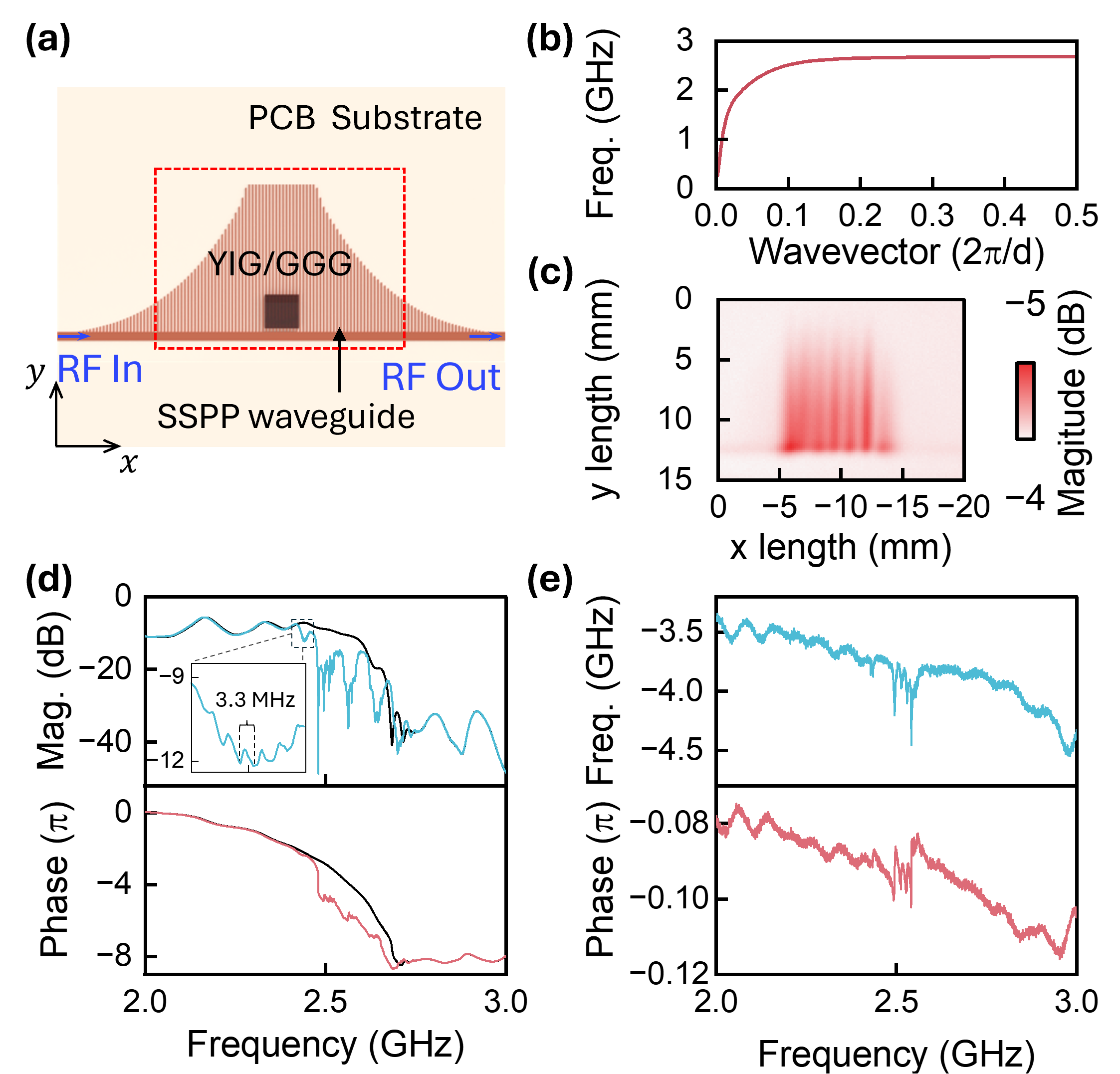}
    \caption{(a) Schematic of the magnon--phonon coupling platform, where a YIG/GGG chip is flip-bonded onto an SSPP waveguide. (b) Simulated dispersion relation of the SSPP waveguide. (c) Measured spatial distribution of the $z$-component of the SSPP mode magnetic field. (d) Measured transmission magnitude and phase of the low-GHz SSPP waveguide with a 3~$\mu$m YIG thin-film device (thin black curves); thick black curves show the background signals without the YIG film. (e) Transmission spectrum of a reference microstrip waveguide with a 3~$\mu$m YIG thin film, exhibiting significantly weaker magnon resonances.}
    \label{fig1}
\end{figure}

The device used to investigate magnon–phonon interactions in the low-GHz regime is illustrated in Fig.~\ref{fig1}(a). The magnonic resonator consists of an YIG thin film grown on a 500 $\mu$m gadolinium gallium garnet (GGG) substrate. Three YIG film thicknesses ($d$) are used: Device 1 (200 nm), Device 2 (3 $\mu$m) and Device 3 (20 $\mu$m). The YIG thin film is biased using an external magnetic field from an electromagnet. The magnonic chip is flip bonded on a spoof surface plasmon polariton (SSPP) waveguide, with the YIG side attached to the waveguide. The SSPP waveguide is a periodically corrugated microstrip, which can drastically enhance the excitation and readout of the magnon signals without sacrificing the bandwidth due to the slow wave nature of the SSPPs. It is fabricated on a Rogers TMM10i printed circuit board (PCB) with a dielectric constant of 9.8, a substrate thickness of 760 $\mu$m, and a copper thickness of 35 $\mu$m thickness. To minimize reflection and interference, the SSPP waveguide is adiabatically tapered on the two ends for better impedance matching with the input microstrip, which is soldered with SMA connectors for transmission measurements. 

The theoretical dispersion curve of the fundamental SSPP mode for this waveguide is plotted in Fig.~\ref{fig1}(b), which is simulated using finite element method (FEM). The corrugation fingers were designed with a length of 3 mm, leading to an asymptotic frequency of around 2.6 GHz. Fig.~\ref{fig1}(c) plots the spatial distribution of the SSPP mode profile near 2.6 GHz, which is measured by scanning an inductive antenna at the waveguide surface in the region shown in the dashed box in Fig.~\ref{fig1}(a). Strong mode confinement is observed near the bottom of the fingers where the YIG chip is located, further enhancing the SSPP--magnon coupling.

Fig.~\ref{fig1}(d) shows the measured transmission spectrum of our SSPP waveguide with YIG Device~2, obtained using a vector network analyzer (VNA) at a power of $-20~\mathrm{dBm}$ to avoid nonlinear effects. When the magnon mode is far detuned, the transmission spectrum (dark blue and red curves) exhibits a clear cutoff near $2.6~\mathrm{GHz}$, with a $30~\mathrm{dB}$ drop in transmission and significant phase accumulation, in good agreement with the simulated dispersion relation. When the magnon modes are tuned close to the cutoff frequency, absorption dips with extinction ratios exceeding $40~\mathrm{dB}$ are observed, enabled by the strong enhancement due to the slow-wave nature of the SSPP mode. For comparison, the same YIG device measured using a conventional microstrip waveguide exhibits much weaker magnon resonances, with extinction ratios of only $0.6~\mathrm{dB}$ [Fig.~\ref{fig1}(e)]. The enhanced coupling between magnons and microwave photons (in the form of the SSPP mode) enables further investigation of magnon--phonon coupling in the low-GHz frequency range. As shown in the close-up view of Fig.~\ref{fig1}(d), periodic oscillations with a period of $3.3~\mathrm{MHz}$ can be clearly resolved on the magnon absorption dip, corresponding to transverse HBAR phonon modes.

\begin{figure}[tb]
    \centering
    \includegraphics[width=\linewidth]{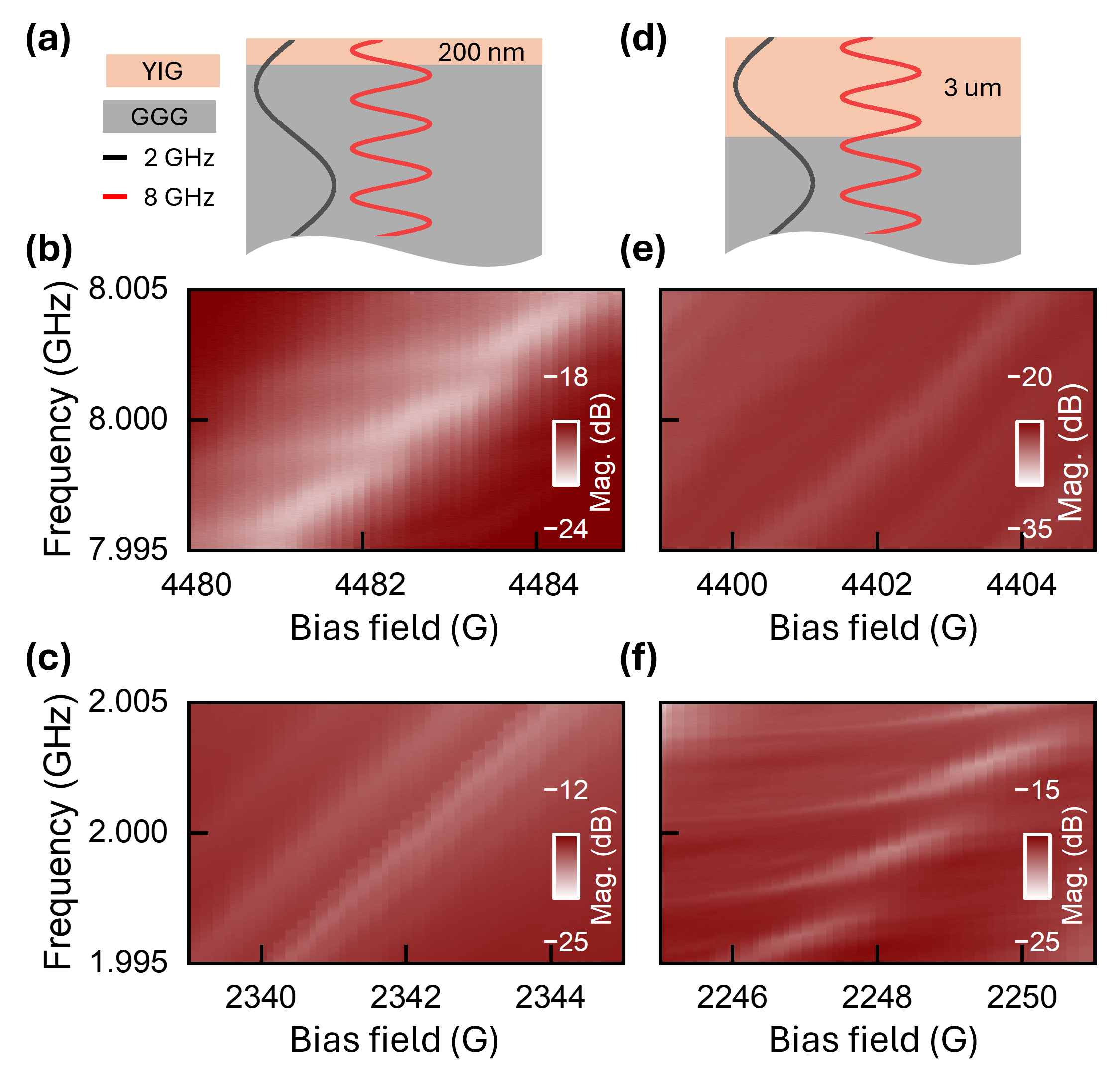}
    \caption{(a), (d) Illustrations of the magnon--phonon coupling regimes in YIG/GGG devices with YIG thicknesses of 200~nm and 3~$\mu$m, respectively. (b), (c) Measured transmission spectra for the 200~nm device using SSPP waveguides operating at 8~GHz and 2~GHz. (e), (f) Measured transmission spectra for the 3~$\mu$m device under the same frequency conditions. The comparison highlights the strong dependence of magnon--phonon coupling on YIG thickness and operating frequency.}
    \label{fig2}
\end{figure}

\section{YIG thickness comparison}

We systematically studied the magnon--phonon coupling across different YIG film thicknesses ($d$) and frequency regimes, with the results summarized in Fig.~\ref{fig2}. We first measured the coupling on a 200-nm YIG thin film (Device~1)---commonly used in previous high-frequency magnomechanical studies---using our low-GHz SSPP waveguide. Our calculation shows that the wavelength of transverse phonons in the YIG/GGG device is approximately 438~nm at 8~GHz, which is comparable to twice the YIG film thickness, as indicated in Fig.~\ref{fig2}(a). This condition corresponds to near-optimal spatial overlap between phonon modes and low-order magnon modes, favoring strong coupling. In contrast, at 2~GHz the phonon wavelength is about 1750~nm which is much larger than $2d$, resulting in a significant mode-volume mismatch and consequently much weaker magnon--phonon coupling.

These effects are evident in the measured transmission spectra at both frequencies, shown as heatmaps in Figs.~\ref{fig2}(b) and (c). When the bias magnetic field is scanned around 4402~G, the magnon mode sweeps across a 10~MHz range near 8~GHz. Periodic features with a spacing of 3.3~MHz are observed, indicating coupling to transverse phonon modes. This period matches the calculated free spectral range (FSR) of HBAR phonons for the given GGG substrate thickness (500~$\mu$m). From these spectra, the magnon--phonon coupling strength is extracted as 0.7~MHz. Although not in the strong-coupling regime, this value corresponds to a cooperativity of $C = 0.6$ when considering the dissipation rates of the magnon mode ($\kappa_m = 2.0$~MHz) and phonon mode ($\kappa_b = 0.4$~MHz), placing the device in the coherent-coupling regime. By contrast, when the same device is measured near 2~GHz, no clear magnon--phonon coupling features are observed, consistent with the prediction above.

The situation changes dramatically for Device~2, which uses a 3~$\mu$m-thick YIG film on the same SSPP waveguide [Fig.~\ref{fig2}(d)]. The increased thickness improves spatial overlap with phonon wavelengths near 2~GHz, enhancing magnon--phonon coupling in the low-GHz regime. Conversely, at 8~GHz the phonon wavelength becomes much smaller than the YIG thickness, leading to reduced coupling due to cancellation effects within a phonon period. This behavior is confirmed by the measured spectra: clear magnon--phonon coupling features appear near 2~GHz but not at 8~GHz. The extracted coupling strength at 2~GHz is approximately 0.9~MHz, corresponding to a cooperativity of $C = 1.0$, both exceeding the values obtained for Device~1 at 8~GHz. This improvement can be attributed to the increased fill factor of the magnon mode (across the YIG thickness) relative to the phonon mode (extending through the entire YIG/GGG substrate) in the thicker YIG film.

\begin{figure}[tb]
    \centering
    \includegraphics[width=\linewidth]{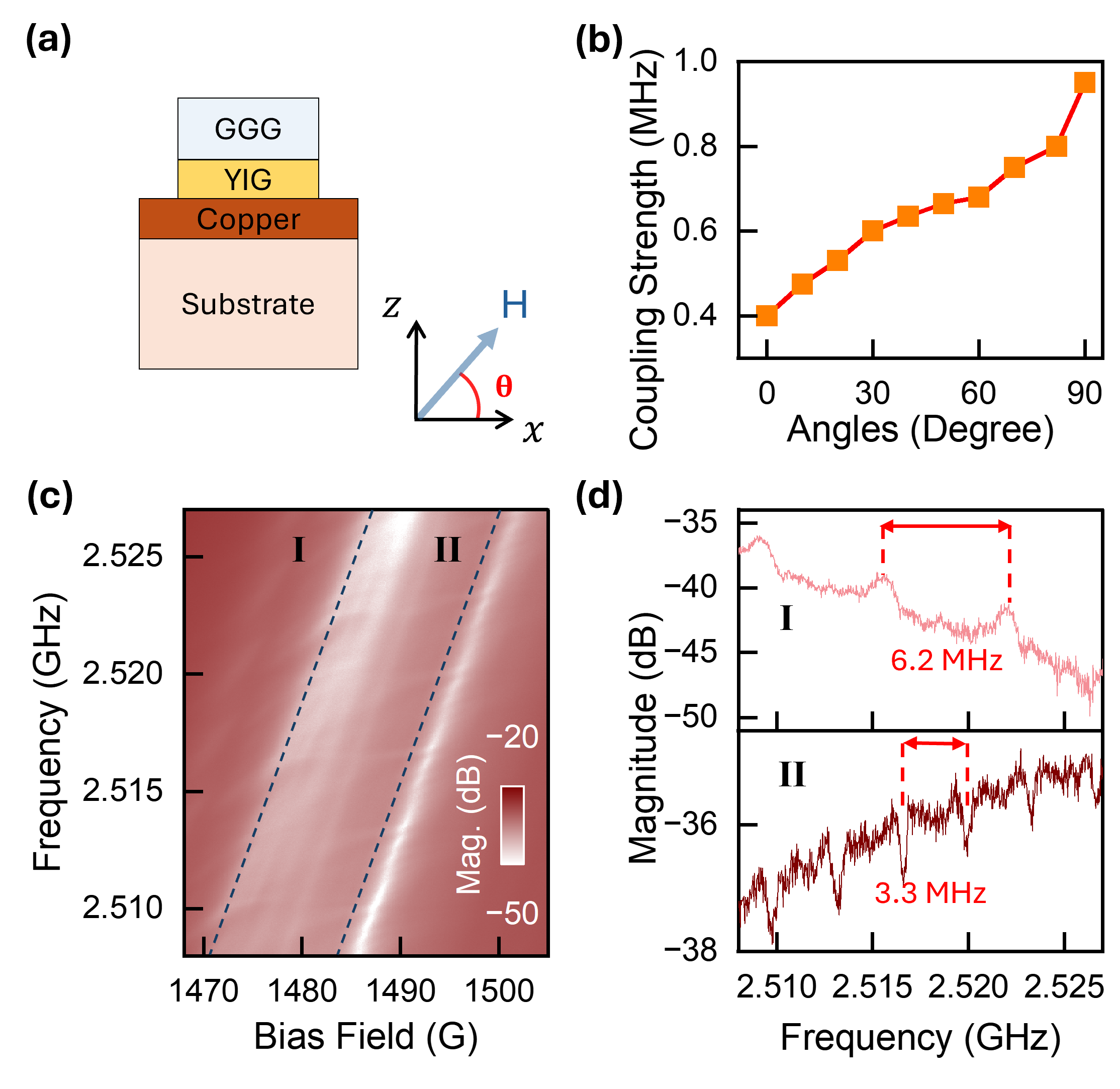}
    \caption{(a) Schematic of the angle-dependent characterization for the 3~$\mu$m YIG thin-film device, where the bias magnetic field $H$ is applied at an angle $\theta$ relative to the $x$-axis. (b) Measured magnon--phonon coupling strength as a function of $\theta$, showing a clear increasing trend with angle. (c) Two-dimensional transmission spectrum at $\theta = 70^\circ$, revealing multiple phonon modes. (d) Spectra corresponding to the two line cuts in (c), exhibiting longitudinal phonon modes (top panel: line cut I, with a uniform spacing of 6.2~MHz) and transverse phonon modes (bottom panel: line cut II, with a uniform spacing of 3.3~MHz), respectively.}
    \label{fig3}
\end{figure}

\section{Angle dependence}

Depending on the direction of the bias magnetic field $H$, different types of magnon modes are supported in the YIG thin film [Fig.~\ref{fig3}(a)]. When the bias field is applied out of plane ($\theta = 90^\circ$), the magnons form forward volume magnetostatic wave (FVMSW) modes. In contrast, for an in-plane bias field ($\theta = 0^\circ$) perpendicular to the SSPP waveguide propagation direction, the magnon modes correspond to magnetostatic surface waves (MSSWs). In our experiment, an automated rotational stage continuously rotates the bias magnetic field in the $y$--$z$ plane, enabling measurements of phonon--magnon interactions under different bias conditions.

As shown in \textcite{2021_PRAppl_Xu_pulseEcho}, transverse phonons dominate the hybridization dynamics with FVMSW magnons, a result attributed to the specific orientations and spatial distributions of the magnonic and phononic modes. As the bias field direction changes, the device deviates from the optimal coupling condition, reducing the coupling strength. This trend is confirmed by the measured coupling strength as a function of bias angle [Fig.~\ref{fig3}(b)], obtained from a 3~$\mu$m YIG thin film coupled to an SSPP waveguide operating near 2.52~GHz. For example, the extracted magnon--phonon coupling strengths at $\theta = 0^\circ, 30^\circ, 60^\circ,$ and $90^\circ$ are approximately 0.4, 0.6, 0.68, and 0.95~MHz, respectively, clearly revealing the angle dependence.

Our measurements at different bias angles also reveal distinct phonon modes. In the YIG/GGG device, both transverse and longitudinal HBAR phonons exist. When the YIG film is magnetized normal to the plane, longitudinal phonons are absent in the magnon spectrum because their interaction with FVMSW magnons is nearly suppressed. However, as the bias field is tilted, the magnon modes are no longer purely FVMSWs, and coupling to longitudinal phonons becomes evident. This effect is illustrated in Fig.~\ref{fig3}(c), measured at $\theta = 70^\circ$, where two sets of phonon modes appear as evenly spaced resonances. The spectra at the two line cuts (I and II) indicated by dashed lines are shown in Fig.~\ref{fig3}(d), clearly displaying the periodic distribution of the two phonon families. These two sets exhibit different free spectral ranges (FSRs): one with a period of 6.2~MHz and the other 3.3~MHz, consistent with the calculated FSRs for longitudinal and transverse modes, respectively.

\section{Multimodes Strong Coupling}

One prominent feature observed in the low-GHz magnomechanical device is the multimode magnon--phonon coupling. As discussed above, achieving reasonable magnon--phonon coupling in this frequency range requires thicker YIG films, which naturally host numerous high-order magnon modes in the form of standing waves along both the thickness and lateral directions. Consequently, the low-GHz magnomechanical device exhibits rich multimode dynamics between these high-order magnon modes and the HBAR phonons.

\begin{figure}[tb]
    \centering
    \includegraphics[width=\linewidth]{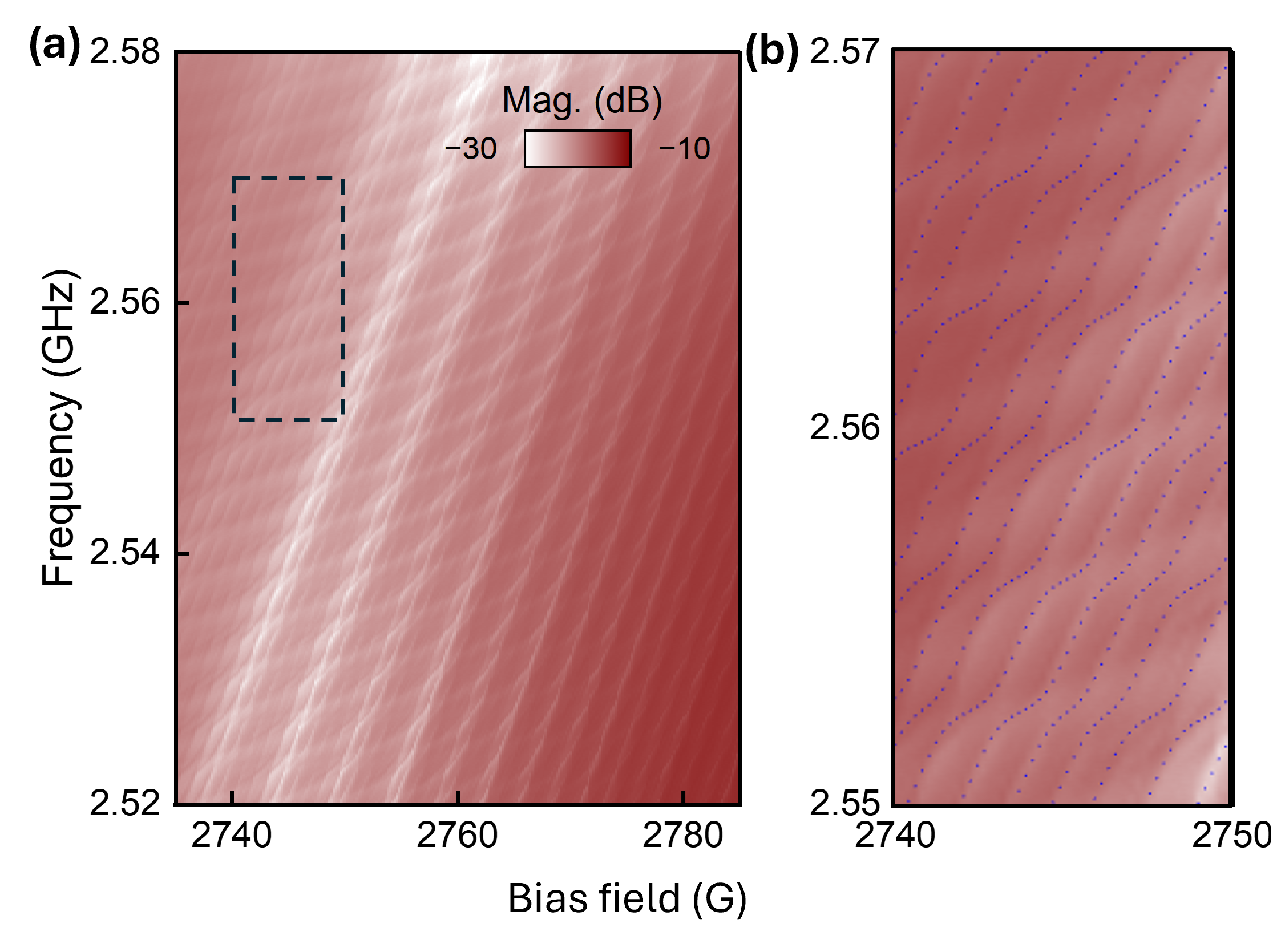}
    \caption{(a) Two-dimensional heatmap of the measured transmission spectrum from a 2.6~GHz SSPP waveguide with a 20~$\mu$m YIG film. (b) Close-up view of the dashed region in (a), with dotted lines indicating theoretically calculated hybrid-mode frequencies.}
    \label{fig4}
\end{figure}

The measured transmission spectrum for a 20~$\mu$m YIG film (Device~3) is shown in Fig.~\ref{fig4}(a), revealing a series of magnon modes forming a mesh-like pattern with clear anticrossings at each mode intersection. The dotted lines in the close-up view [Fig.~\ref{fig4}(b)] represent theoretically calculated hybrid-mode frequencies, which agree well with the experimental results. From these calculations, the magnon--phonon coupling strength in Device~3 is extracted to be approximately 0.6~MHz. Notably, unlike the nearly evenly spaced HBAR phonon modes, the magnon modes exhibit nonuniform frequency spacing due to their strong dispersion.

Such coherent coupling between multiple high-order magnon modes and HBAR phonons enables information exchange in a high-dimensional state space, significantly enriching the functionality of hybrid magnonic systems. Multimode interactions provide access to complex interference effects, mode-selective dynamics, and enhanced control over energy flow, which are not achievable in single-mode systems. This opens opportunities for implementing advanced protocols such as parallel information processing, frequency-multiplexed transduction, and entanglement generation across multiple modes. 

In particular, when combined with dynamic control techniques such as Floquet driving \cite{Xu2020FloquetCavityElectromagnonics,Pishehvar2025OnDemandDarkMode,Pishehvar2025ResonanceEnhancedFloquet}, selective and on-demand coupling between individual modes becomes feasible. Floquet engineering can modulate the effective interaction landscape, enabling programmable connectivity among magnon and phonon modes. Such capabilities are highly desirable for hardware-efficient quantum information processing, where multimode architectures can reduce resource overhead and support scalable implementations of quantum networks and quantum memories. Beyond quantum applications, these multimode hybrid systems also offer a promising platform for studying collective phenomena, non-Hermitian physics, and topological effects in strongly coupled bosonic systems.

\section{Conclusion}
In this work, we experimentally investigate strong magnon–phonon coupling in YIG thin films within the low-GHz regime—a frequency range of significant practical interest but has only received limited prior attention in the study of magnomechanics. By employing slow-wave waveguides SSPPs, we achieve substantial enhancement of magnon–microwave photon coupling near 2~GHz, enabling systematic exploration of thin-film magnomechanical interactions in this domain. Our measurements reveal the dependence of coupling strength on YIG thickness and magnetization orientation, uncovering interactions between magnons and both transverse and longitudinal phonon modes. Furthermore, we observe multimode coupling between higher-order magnon modes and HBAR phonons in thicker films, opening avenues for complex mode-selective dynamics. Collectively, these results establish a versatile platform for low-GHz magnomechanical devices and underscore their potential for coherent microwave-to-acoustic transduction, multimode quantum processing, and hardware-efficient hybrid quantum systems.

\begin{acknowledgments}
Y.J., Z.Y. and X.Z. acknowledge support from NSF (2337713). 
\end{acknowledgments}

\section*{AUTHOR DECLARATIONS}
\textbf{Conflict of Interest}

The authors have no conflicts to disclose.\\

\textbf{DATA AVAILABILITY}

The data that support the findings of this study are available
from the corresponding author upon reasonable request.

\vspace{12pt}

\bibliography{REFM}

\end{document}